\documentstyle[11pt,newpasp,twoside,epsf]{article}
\def\edcomment#1{\iffalse\marginpar{\raggedright\sl#1\/}\else\relax\fi}
\reversemarginpar
\def\ecs{erg~cm$^{-2}$s$^{-1}$}
\def\ufl{erg~cm$^{-2}$}

\begin{document}
\title{X/$\gamma$-ray measurements of the faint GRB~020321}
 \author{J.J.M. in 't Zand, L. Kuiper, J. Heise}
 \affil{Space Research Organization Netherlands \&
 University Utrecht}
 \author{L. Amati, E. Costa, M. Feroci, F. Frontera, G. Gandolfi, L. Nicastro, L. Piro}
 \affil{CNR Istituto Astrofisica Spaziale e Fisica Cosmica, Italy}
 \author{P. Rodriguez-Pascual, M. Santos-Lleo, N. Schartel}
 \affil{XMM-Newton Operations Centre, European Space Agency, Madrid, Spain}

\begin{abstract}
GRB~020321 is a faint GRB that received wide follow-up attention in
X-rays (BeppoSAX/NFI, Chandra/ACIS-S, XMM-Newton), radio (ATCA) and
optical (ESO, HST). We identify a weak X-ray afterglow by a combined
study of the Chandra and XMM-Newton observations. Its decay index of
1.2 is typical for GRB afterglows. Inside the 4\arcsec\ accurate error
box there is a weak optical counterpart candidate with a much
shallower decay index.
\end{abstract}

\section{Introduction}

GRB~020321 was detected by the BeppoSAX Wide Field Camera unit 1 (WFC;
Jager et al. 1997) and Gamma-Ray Burst Monitor (GRBM; Costa et
al. 1998) at 4:20:40 U.T. (Gandolfi 2002a). The WFC-determined
position, 7\deg\ from the celestial south pole, was disseminated 6
hours after the burst. Due to a non-optimum attitude solution the
positional accuracy was limited to 5$^\prime$ (99\% confidence). The
burst was followed up with major observatories in X-rays (Chandra
[Fox 2002] and XMM-Newton [public TOO]), optical (ESO, HST), and radio
(ATCA; Wieringa et al. 2002).  We here summarize the X/$\gamma$-ray
measurements.

\section{Prompt emission}

\begin{figure}
\plotone{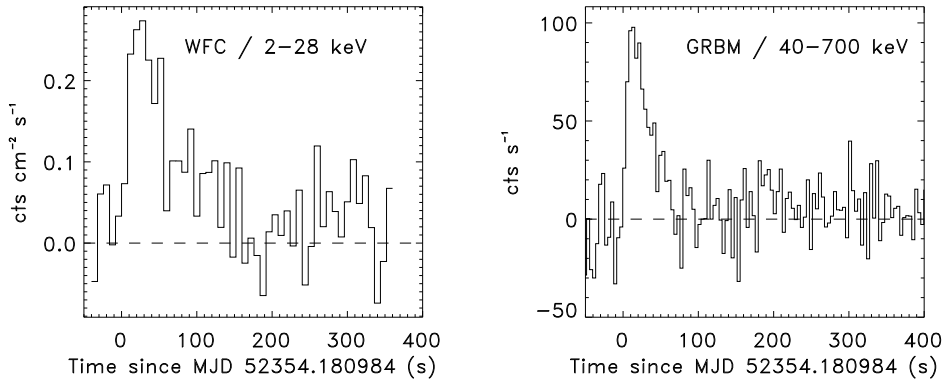}
\caption[]{GRB~020321 light curves in X-rays (left) and
$\gamma$-rays (right).\label{figp}}
\end{figure}

Fig.~\ref{figp} shows the time profiles from the WFC and
GRBM data. The peak intensities are among the lowest of the
bursts and flashes seen with the WFC. The $\gamma$-ray duration is
typical at about 70 sec.  Due to the faintness, the GRBM was not
triggered into a high time-resolution data acquisition mode. A
hard-to-soft evolution can be discerned from comparing X-ray and
$\gamma$-ray data.  The complete WFC+GRBM spectrum is consistent with
a 'Band' spectrum (Band et al. 1993; $\chi^2_\nu=0.52$ for 12 dof)
with parameters $\alpha=-0.95\pm0.25$, $\beta=-2.01\pm0.68$ and
$E_{\rm p}=144\pm130$~keV. The 2-10 to 50-300 keV fluence ratio is
0.07 which is typical for GRBs (cf, Heise et al. 2001). Further
details are listed in Table~\ref{tab}.

\begin{center}
\begin{table}[b]
\caption[]{Characteristics of GRB~020321.\label{tab}}
\begin{tabular}{lll}
\tableline
Parameter & Band & Value \\
\tableline
Duration & 2--28 keV  & $\approx$140 s \\
         & 40--700 & $\approx$70 s \\
Peak flux& 2--10   &  $4\times10^{-9}$~\ecs \\
         & 50--300 & $5\times10^{-8}$~\ecs \\
         & 40--700 & $1\times10^{-7}$~\ecs \\
Fluence  & 2--10   & $9\times10^{-8}$~\ufl \\
         & 50--300 & $2\times10^{-6}$~\ufl \\
         & 40--700 & $3\times10^{-6}$~\ufl \\
Isotropic energy output for $z=1$ & 20-2000  & $6\times10^{52}$~erg \\
\tableline\tableline
\end{tabular}
\end{table}
\end{center}

\section{X-ray afterglow}

\begin{figure}
\plotone{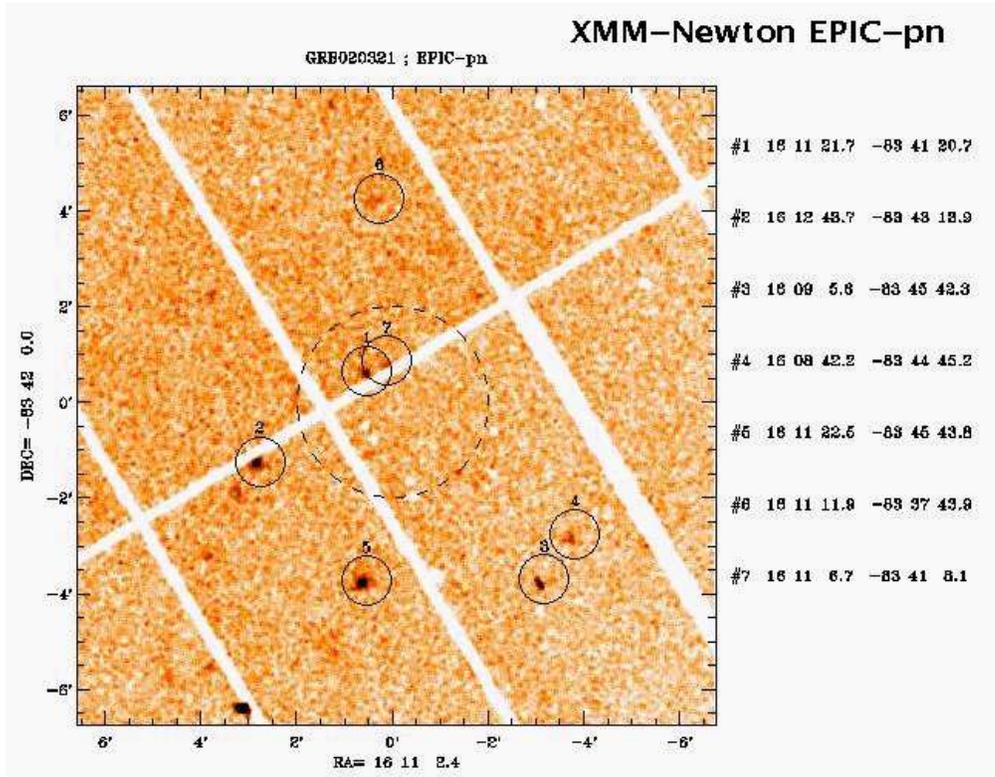}\\
\plotone{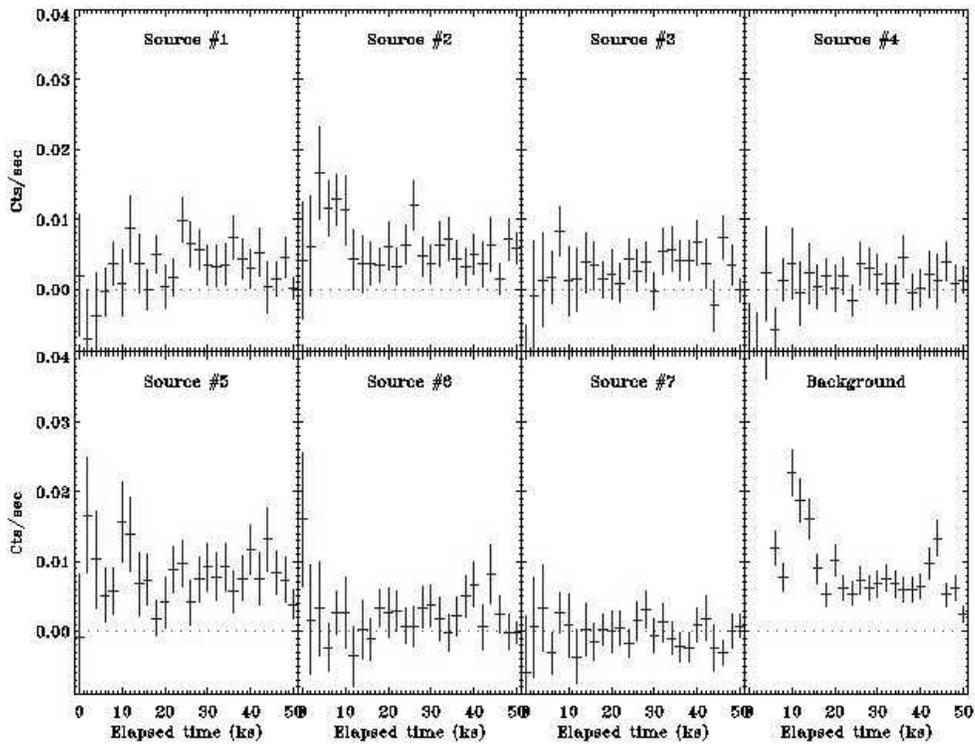}
\caption[]{EPIC PN image (top) and light curves based on preliminary
net count rates of all 7 detected sources (bottom). The large circle
in the image is the error circle of the initially reported but
unconfirmed X-ray afterglow (Gandolfi 2002b; In 't Zand et al. 2002).
\label{figxmm}}
\end{figure}

\begin{figure}
\plotfiddle{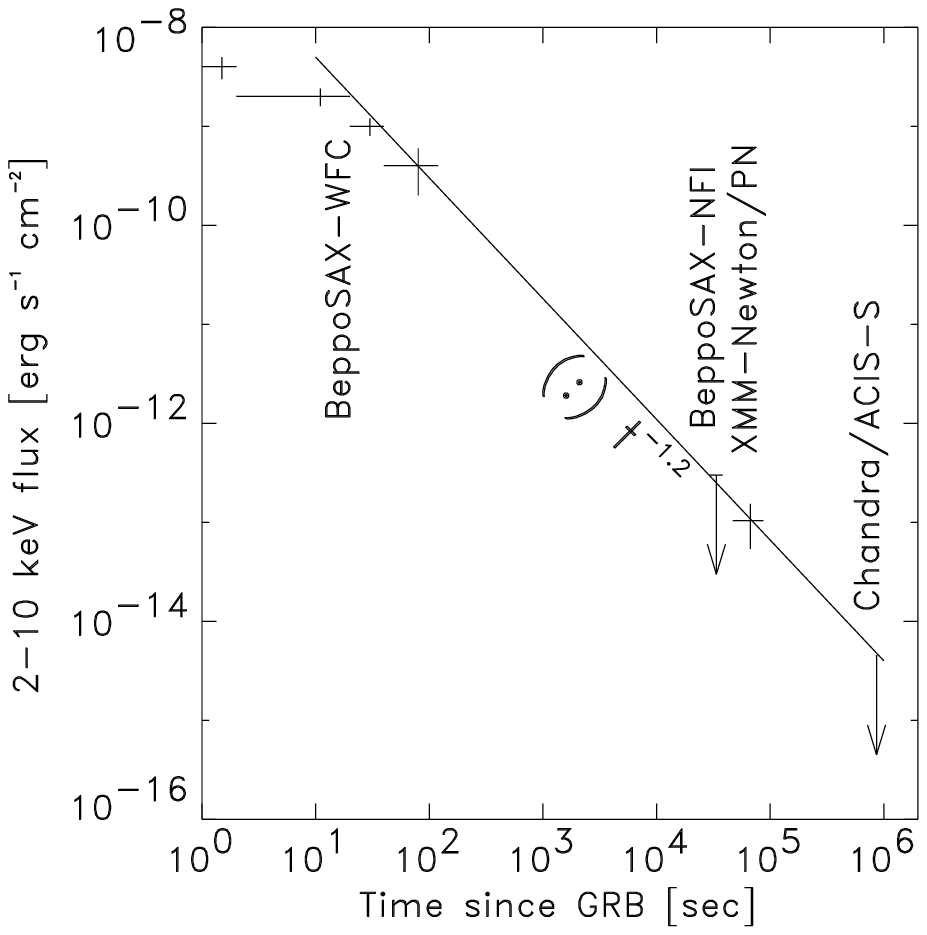}{5.5cm}{0.}{80.0}{70.0}{-144}{-67}
\caption[]{X-ray light curve. Arrows
indicate upper limits. The XMM-Newton point is based on the latter
40~ksec of the observation.\label{figlonglc}}
\end{figure}

The BeppoSAX Narrow Field Instruments followed up for only a limited
time, from 8.1 to 10.6 hrs after the burst.  The exposure time for the
MECS instrument is 6133~s. Although there was an initial report of an
afterglow detection (Gandolfi 2002b), a refined analysis could not
confirm this (In~'t~Zand et al. 2002). No source was detected above an
upper limit of $3\times10^{-13}$~erg~s$^{-1}$cm$^{-2}$ (2-10 keV; for
a Crab spectrum).  XMM-Newton followed up from 10.3 to 24.2 hrs after
the burst, for an exposure time of 50 ksec. Seven sources are obvious
in the PN-image (see Fig.~\ref{figxmm}).  Chandra followed up 10.0
days after the burst for 20 ksec with ACIS-S and no grating.  Seven
sources are apparent (Fox 2002). The S-3 chip only covers 3 of the XMM
sources. Two of these are detected, but XMM-Newton 'source 2' is not,
thus it is a good afterglow candidate. The position as determined
with XMM-Newton has an error circle with radius 4$^{\prime\prime}$.
Figure~\ref{figlonglc} presents the 2-10 keV flux time history of the
prompt emission and 'source 2'. The combined data are consistent with
a decay index of 1.2.

\section{Discussion}

It was only after 18~d that a promising X-ray afterglow candidate
could be identified, with a position 3\farcm 0 off the initially
reported afterglow.  This is probably the reason why some follow up
was not successful. Nevertheless, observations with the 3.6~m ESO
telescope at La Silla covered source 2 and show two extended sources
in the error box of the X-ray afterglow (Salamanca et al. 2002). One
of them shows clear signs of decay in $R$, with $R=22.84$ at 1.1 days
after the burst. The decay index is $\approx0.2$. Salamanca et
al. propose that the shallow index may be due to an optical transient
superposed on a host galaxy.


\begin{references}
Band, D., et al. 1993, \apj, 413, 281

Costa, E., et al. 1998, Adv. Sp. Sc, 22, (7)1129

Fox, D. 2002, GCN 1342

Gandolfi, G. 2002a and b, GCN 1281 and 1285 respectively

In 't Zand, J.J.M., et al. 2002, GCN 1348

Heise, J., et al. 2001, in 2nd Rome GRB workshop, GRBs in the Afterglow Era,
    eds. E. Costa, F. Frontera and J. Hjorth (Berlin: Springer), 16

Jager, R., et al. 1997, \aaps, 125, 557

Salamanca, I., et al. 2002, GCN 1385

Wieringa, M., et al. 2002, GCN 1308
\end{references}
\end{document}